# Understanding the role of transport velocity in biomotor-powered microtubule spool assembly

Amanda J. Tan,[a] Dail E. Chapman,[b] Linda S. Hirst,[a] and Jing Xu[a†]

[a] Physics, University of California, Merced, CA 95343, USA.

[b] Developmental and Cell Biology, University of California, Irvine, CA 92697, USA.

[†] Correspondence: jxu8@ucmerced.edu

**Abstract**

We examined the sensitivity of microtubule spools to transport velocity. Perhaps surprisingly, we determined that the steady-state number and size of spools remained constant over a seven-fold range of velocities. Our data on the kinetics of spool assembly further suggest that the main mechanisms underlying spool growth vary during assembly.

**Introduction**

The emerging field of active matter, in which the elements of a material consume energy to move relative to each other, has recently focused on biological examples in cell-free systems.[1, 2] An example of an active behavior can be seen in gliding assays in which biopolymers are transported, or "glide", across a biomotor-decorated surface.[1] Biomotors are protein machines that convert chemical energy stored in adenosine triphosphate (ATP) into mechanical motion to transport materials in cells.[3] Biopolymers such as microtubules are molecular roads for biomotor-powered transport in cells.[4] Biomotors and their biopolymer roads are fundamental building blocks of eukaryotic life; they are critical for maintaining cell function, as well as for controlling cell shape and cell movement.[3, 5, 6] There is increasing interest in using these biological building blocks to assemble higher-order structures for bioengineering applications.

Spooling is an example of biomotor-propelled active assembly in which linear microtubules form micron-sized, ring-shaped structures, called "spools".[7, 8] To generate spools, biotin-functionalized microtubules are used in standard gliding assays, and additional streptavidin molecules are added to introduce "sticky" interactions between microtubules (Fig. 1, Mov. S1, and Experimental section in the Supplementary Information). Biomotor-powered microtubule spools represent a promising model for engineering biological transducers, as they convert chemical energy input into mechanical energy (to bend microtubules) as well as kinetic energy (to sustain spool rotation).

To date, several factors have been found to impact spool assembly, including motor density[9], microtubule length and rigidity[10], the density and interaction strength between microtubules,[11, 12] flow cell material,[13-15] and step-wise assembly conditions.[16] The role of microtubule transport velocity in spool assembly, however, has remained unclear.

Previous studies suggested a role of transport velocity in spool assembly.[9, 12, 17, 18] First, transport velocity appeared to influence the number of assembled spools, as fewer spools formed at lower transport velocity.[12] However, this effect may reflect a lower kinetic rate of spool assembly rather than a smaller probability of spool formation, as spool assembly was characterized at a single time point (15 min) after initiating spool assembly. Second, transport velocity appeared to influence the morphology of microtubule assembly.[17] At lower transport velocities, a transient state in which microtubules form long linear bundles, in addition to spools, was observed.[17] Whether/how such changes in assembly morphology impact the properties of spools assembled at the lower transport velocity has remained unclear. Finally, transport velocity can impact the number of biomotors simultaneously available to propel microtubules.[18] Because this motor number can affect both the size and the number of assembled spools,[9] we hypothesized that spool size and/or number is influenced by transport velocity. To test this hypothesis, we examined the role of microtubule transport velocity in spool assembly, both in terms of assembly kinetics and the properties of spools at steady state (Figs. 2-3).

**Results and Discussions**

**Transport velocity influences the rate of spool assembly**

We first examined the kinetics of spool assembly for different transport velocities (Fig. 2). We used the major microtubule-based biomotor, kinesin-1, to power spool assembly. In order to tune microtubule transport velocity, we varied the concentration of ATP in our flow cell as in previous studies,[12, 19-21] and carried out spooling experiments under otherwise identical conditions (Fig. S1). In order to keep the ATP concentration (and thus the transport velocity) constant throughout each experiment, we used an ATP regeneration system as previously described.[21-23] For each transport velocity, we measured the number of spools assembled as a function of observation time for up to 2 h after the introduction of ATP into the flow cell (Fig. 2a). Our observation time was limited to 2 h because we detected substantial deterioration/breakage of microtubules in our field of view at ~2 h (data not shown). Such deterioration of gliding microtubules is consistent with the progressive loss of microtubules ("molecular wear") reported previously[24] for similar experimental systems.

We found that a non-zero transport velocity is necessary for the active assembly of microtubules into spools (Fig. 2a). Before ATP was introduced into the flow cell, the microtubules remained static and isolated in our field of view (Fig. 2a). This observation is perhaps not surprising because biomotors rely on ATP hydrolysis to propel microtubules[19, 20]; some level of microtubule gliding should be necessary to reduce the interaction distance between them and to allow spool assembly. Although the mechanisms underlying spool formation are not fully understood, all current models[13, 25] implicitly require that the microtubules glide at a finite velocity. Consistent with this notion, once microtubule gliding was initiated (via the introduction of ATP), we observed clear spool assembly (Fig. 2a).

We found that the kinetics of spool formation differed substantially between transport velocities (Fig. 2b). At saturating ATP and thus maximum transport velocity (Fig. 2b, iii), we detected a nonlinear dependence of spool number on assembly time. The number of assembled spools reached a maximum at ~15 min, before reducing by ~20% over the next ~40 min and remaining approximately constant thereafter (Fig. 2b, iii). This non-linear behavior is consistent with a previously reported metastable stage[26] and indicates that spools can disassemble to some extent before assembly reaches a steady state. At the lower velocities tested (31±1 nm/s and 190±2 nm/s, Fig. 2b), we did not observe any clear metastable stage. Instead, the kinetics of spool assembly were well characterized by the asymptotic function, $A \cdot (1-e^{-t/\tau})$ (Fig. 2b, i-ii). Independent of the metastable stage, our data demonstrate that the rate of spool assembly depends strongly on microtubule transport velocity, with the best-fit time constant ($\tau$, Fig. 2b) for spool number increasing with increasing transport velocity ($\tau$=29±10 min in Fig. 2b, i vs. $\tau$=3±2 min in Fig. 2b, iii). Thus, the dependence of spool number on transport velocity reported previously[12] reflects at least in part the slower kinetic rates of spool assembly at lower transport velocities.

**Transport velocity does not influence spool density at steady state**

We did not observe any significant influence of transport velocity on the steady-state number of assembled spools (Fig. 2b-c). The asymptotic numbers of spools assembled at the lower velocities (15~18 spools, Fig. 2b, i-ii) are in good agreement with the steady-state number of spools assembled at the maximum transport velocity (~16 spools, Fig. 2b, iii). To account for the possibility that spool assembly did not reach steady state (for example, Fig. 2b, i), we examined how transport velocity influenced the number of spools assembled at the same time point (Fig. 2c), and how this dependence varied as a function of time (Fig. 2c). We found that the impact of transport velocity on spool number varied substantially with assembly time. Within the first ~40 min of assembly, the number of assembled spools increased significantly with increasing transport velocity (for example, 15 min, Fig. 2c). This observation is in good agreement with a previous report by Liu et al.[12] However, the influence of transport velocity on spool number became less pronounced with increasing assembly time (for example, 60 min, Fig. 2c). For times ≥60 min and within the statistical power of our experiments, we did not observe any significant dependence of spool number on transport velocity (Fig. 2c).

Taken together, our data indicate that the steady-state number of spools is not influenced by transport velocity. In other words, microtubules need to be propelled to be within an interaction distance from each other, but the speed at which they are propelled is not critical and does not impact their eventual assembly into higher-order structures (spools).

**Transport velocity influences spool size during initial assembly**

We next examined how spool circumference varied as a function of assembly time (Fig. 3). Perhaps surprisingly, for all velocities tested, spool size increased substantially with increasing time. For example, at a velocity of 190±2 nm/s, the mean circumference of assembled spools at 90 min was ~2.8-fold larger than that measured at 5 min (17±1 μm at 90 min vs. 6±1 μm at 5 min, Fig. 3a, ii).

What underlies the observed increase in spool size with time (Fig. 3a)? The circumference of individual spools may relax/increase over time, perhaps via relative shearing of the microtubules against each other after they are assembled into spools. Such relaxation would highlight the dynamic nature of microtubule assembly, as suggested by the presence of a metastable stage at the fastest transport velocity (~15 min in Fig. 2b, iii; reported previously[26]). However, we did not detect any substantial difference in spool circumference over the same time course (~15 min vs. >60 min, Fig. 3a, iii). Thus, our data do not directly support this relaxation model. Instead, our data are consistent with a model in which spools can disassemble but spool size is predominantly determined upon nucleation. This model is consistent with the recent finding that spool size is strongly related to the specific mechanism of assembly.[13]

We speculate that a smaller initial spool size reflects the close proximity of microtubules at the beginning of assembly (0 min, Fig. 2a): spool size is limited by the presence of microtubules surrounding the spools. This confinement effect is proposed to underlie spool assembly,[13, 17, 25] and was previously demonstrated to induce a transient loop of "non-sticky" microtubules (lacking biotin-streptavidin-mediated adhesive interactions) during gliding.[23] In this scenario, because the microtubules can bundle and spool at faster rates at higher transport velocities (Fig. 2), we expect the magnitude of the confinement effect to decrease and the spool size to increase with increasing transport velocity. This prediction is consistent with our experimental data, as we observed a ~3-fold increase in initial spool size over the range of transport velocities tested (3±1 μm in Fig. 3a, i vs. 9±1 μm in Fig. 3a, iii).

**Transport velocity does not influence spool size at steady state**

We determined that, with increasing assembly time, the average circumference of spools approached similar asymptotic values for all transport velocities tested (14~18 μm, Fig. 3a). This finding is consistent with our speculation of a confinement effect during the initial stage of spool assembly. Because the number of assembled spools increases over time (Fig. 2b-c), most spools are assembled at a later time and are thus minimally confined by the spatial proximity between neighboring microtubules. Taken together, our data suggest that the confinement effect is an important factor for initiating spool assembly, and that the mechanisms underlying spool assembly can change with increasing assembly time (likely due to changing spatial proximity of microtubules in the assay).

We did not detect a substantial or consistent effect of transport velocity on spool circumference at any single assembly time point, for times ≥30 min (Fig. 3b). Thus, within the statistical power of our experiments, the steady-state size of spools is not substantially influenced by transport velocity.

Taken together, our findings suggest that the maximum force produced by a group of biomotors (kinesins) is not sensitive to transport velocity. Biomotors must exert force to bend the microtubule from its linear form to the curved structure in spools. Because the microtubule is highly rigid,[4, 27, 28] the maximum force that biomotors exert to bend microtubules critically determines the size of microtubule spools, regardless of the propelling rate. Since we did not detect any sensitivity of spool size to transport velocity (Figs. 3), the maximum force exerted by the biomotors to propel microtubules must remain approximately constant over the ~7-fold range of transport velocities tested here (31-222 nm/s). This finding is somewhat surprising given a previous report that the number of kinesins simultaneously bound to the microtubule is inversely tuned by transport velocity,[18] and given other reports that the force propelling each microtubule depends on the number of biomotors present.[29, 30] We speculate that transport velocity influences the relative probability of maximal versus intermediate force production by a group of kinesin biomotors. We are working on future optical-trapping studies to examine this potential effect of transport velocity on force production by groups of biomotors.

**Conclusions**

Here we examined the role of transport velocity in the active assembly of microtubule spools, using a cell-free system in which microtubules were propelled by biomotors (kinesins). We found that transport velocity influences the kinetics of spool assembly, but not the steady-state properties of assembled spools. Specifically, transport velocity influences the rate of spool assembly and the size of spools during initial assembly, but does not alter the number or size of assembled spools at steady state. Our data suggest that the confinement effect is an important factor in initiating spool assembly, and that the main mechanisms underlying spool assembly may vary during assembly.

**Acknowledgements**

We thank Henry Hess (Columbia University), David Sivak (Simon Fraser University), and Uri Raviv (Hebrew University of Jerusalem) for helpful discussions. We thank Tiffany J. Vora (Bayana Science) for helpful manuscript editing. We thank the reviewers for their thoughtful suggestions.

This work was supported by the UC Merced Academic Senate Committee on Research Award and by a UC Merced Health Science Research Institute seed grant.

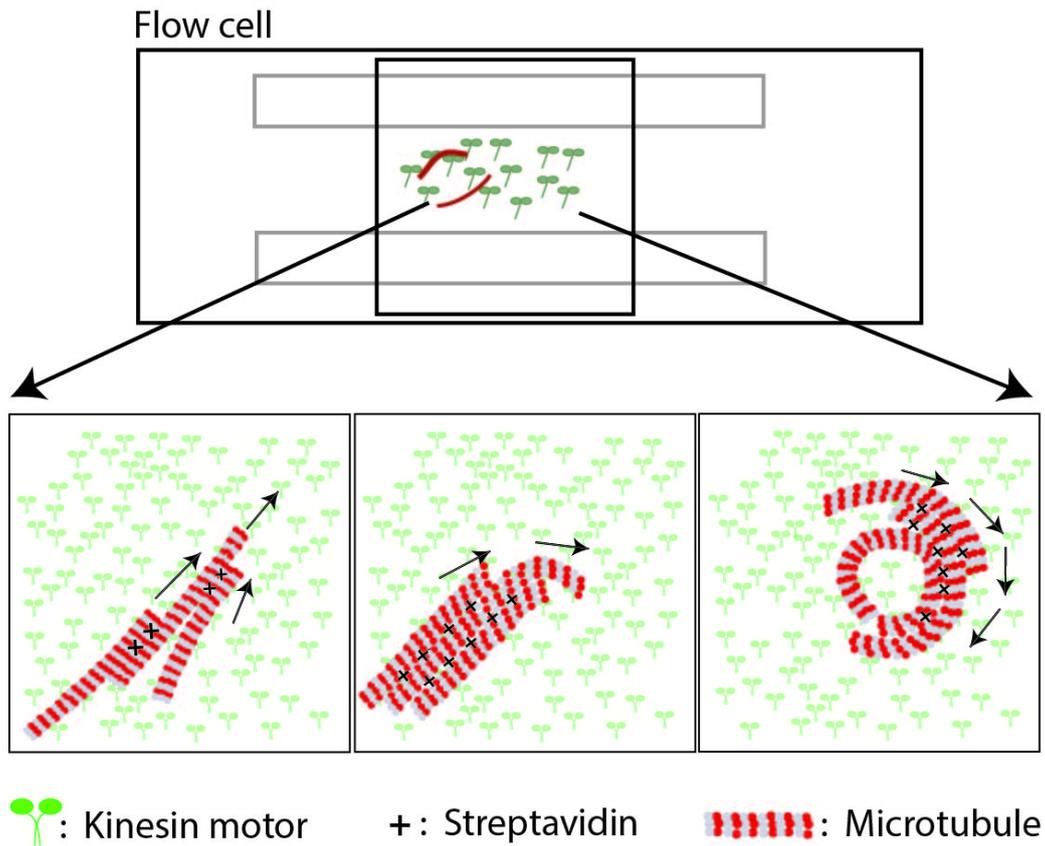

**Figure 1.** Schematic of microtubule spooling experiments (not to scale). Biomotors such as kinesins transport microtubules across the flow-cell surface. Streptavidin was introduced to act as a nano-"adhesive" between microtubules, as previously described.[7,8]

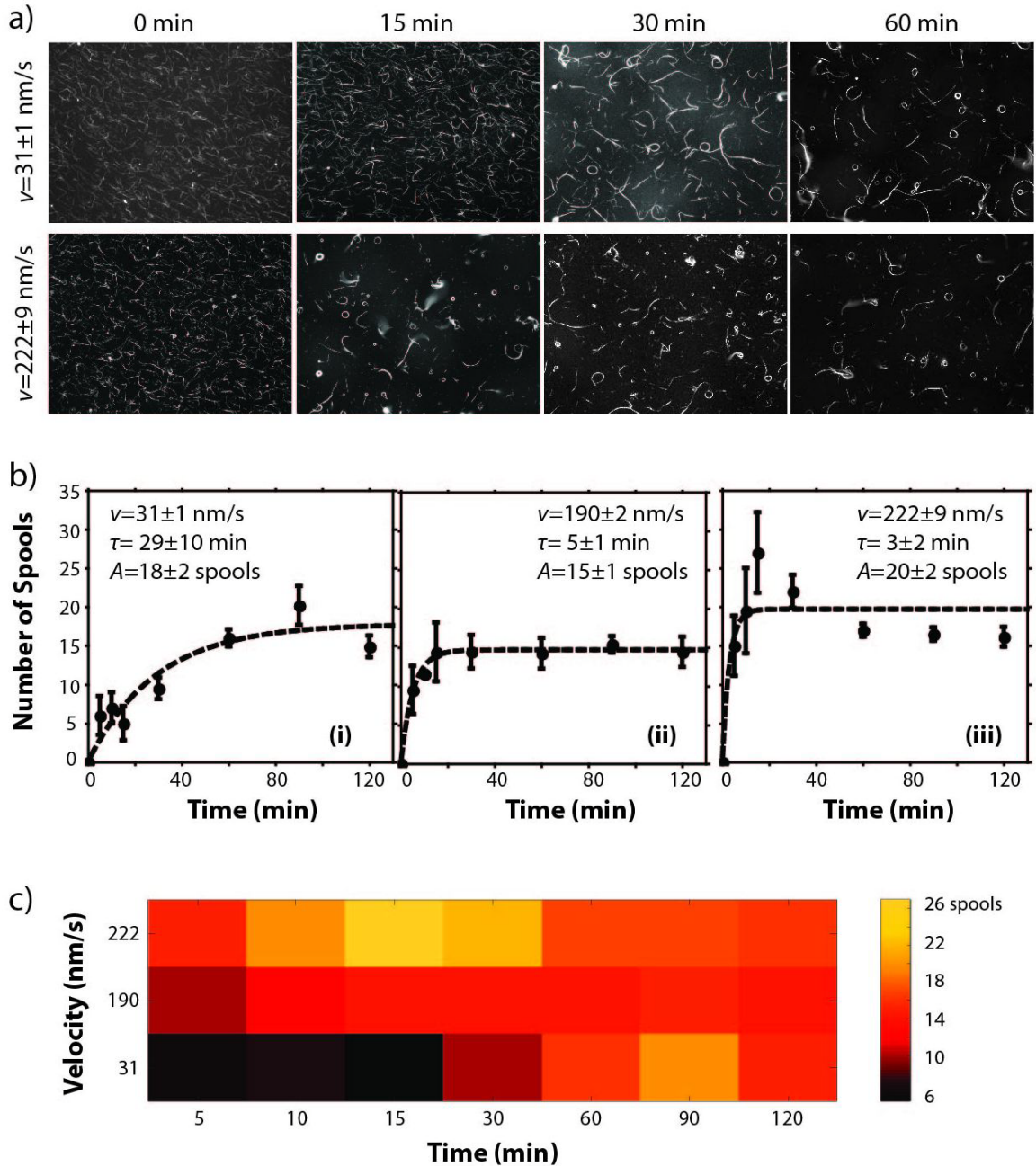

**Figure 2.** Kinetics of spool assembly at distinct transport velocities. (a) Representative images taken at $v=31\pm1$ nm/s (0.05 mM ATP) and $v=222\pm9$ nm/s (1.0 mM ATP). (b) The average number of spools assembled within our field of view, measured as a function of time for three transport velocities. Error bars, standard error. Dashed line, best fit to the asymptotic function $A\cdot(1-e^{-t/\tau})$. $\tau$, best-fit time constant. $A$, best-fit asymptotic number of assembled spools. (c) Heat map of spool number as a function of time and transport velocity.

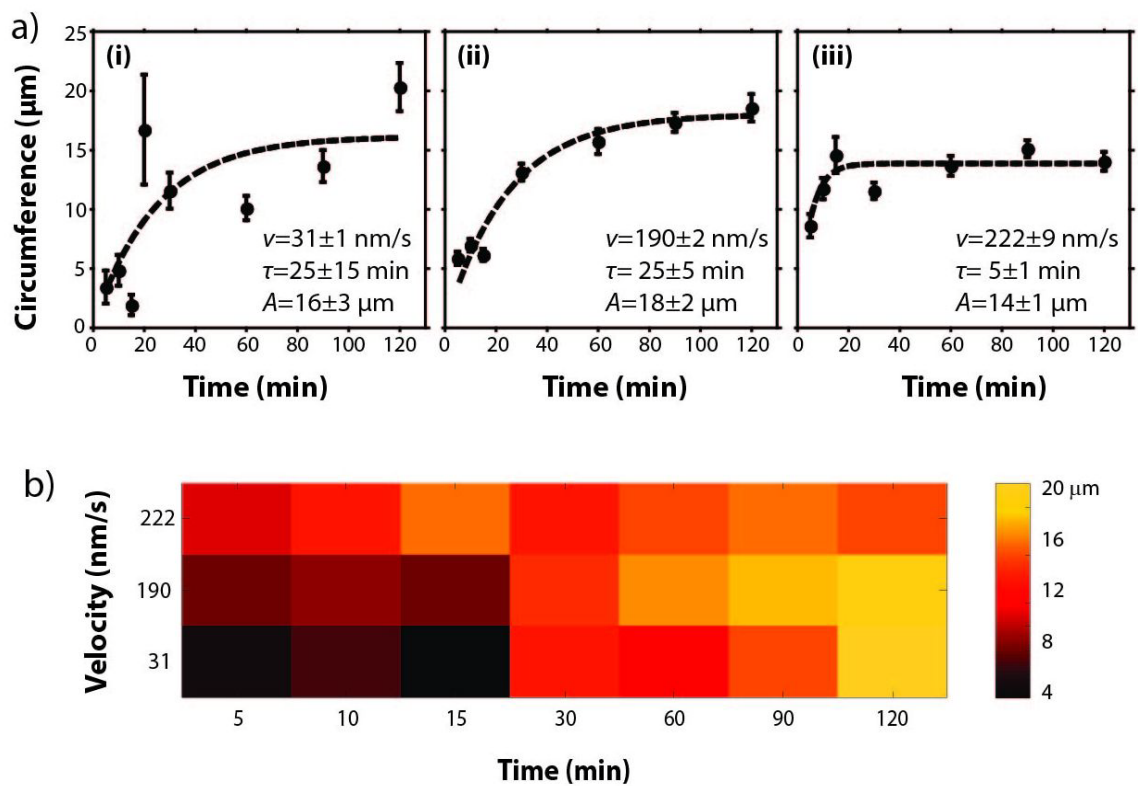

**Figure 3.** Effect of transport velocity on spool circumference. (a) Spool circumference as a function of assembly time for three transport velocities. Error bars, standard error. Dashed line, best fit to the asymptotic function $A \cdot (1-e^{-t/\tau})$. $\tau$, best-fit time constant. $A$, best-fit asymptotic value for spool circumference. (b) Heat map of spool circumference as a function of time and transport velocity.

# SUPPLEMENTAL INFORMATION

## Understanding the role of transport velocity in biomotor-powered microtubule spool assembly

Amanda J. Tan[a], Dail E. Chapman[b], Linda S. Hirst[a], and Jing Xu[a,*]

[a]Physics, University of California, Merced, CA 95343, USA

[b]Developmental and Cell Biology, University of California, Irvine, CA 92697, USA

***Correspondence:** Jing Xu (jxu8@ucmerced.edu)

## TABLE OF CONTENTS

# Experimental

**Proteins and reagents**

Biotin-labeled tubulin (T333P) and rhodamine-labeled tubulin (TL590M) were purchased from Cytoskeleton Inc. (Denver, CO, USA). Streptavidin-fluorescein isothiocyanate conjugate (21224) was purchased from Thermo Fisher Scientific (Waltham, MA, USA). Chemicals were purchased from Sigma Aldrich (St. Louis, MO, USA). Recombinant kinesin consisting of the first 560 amino acids of human kinesin-1 was prepared using plasmid pET17_K560_GFP_His[1] (Addgene, Cambridge, MA, USA). Kinesin was bacterially expressed and Ni-NTA purified as previously described.[2]

**Microtubule preparation**

Microtubules were rhodamine-labeled for fluorescence imaging and biotin-labeled to mediate streptavidin-based lateral interactions between microtubules. Dual-labeled microtubules were prepared by incubating tubulin mix (6:1 biotinylated tubulin to rhodamine-labeled tubulin, 0.47 mg/mL in PEM80 buffer (80 mM PIPES, 1 mM ethylene glycol bis(β-aminoethyl ether), 1 mM $MgSO_4$, pH 6.9)) supplemented with 20 μM taxol and 2 mM GTP for 1 h at 37 °C. Microtubules were kept at room temperature in a dark box and used within four days of preparation.

**Microtubule spooling experiments**

Microtubule spooling experiments were carried out in flow cells in vitro. Flow cells were constructed using a coverslip (22 mm x 22 mm) and a microscope slide, which were held together with double-sided tape (Fig. 1a). Both the coverslip and the microscope slide were biologically clean (accomplished via sequential washing with acetone and ethanol). The volume of each flow cell was approximately 10 μL.

Kinesin solution (100 nM in PEM80 buffer supplemented with 20 mM dithiothreitol) was flowed into the flow cell to bind to the glass surface (5 min). Excess/unbound kinesin was washed away and the flow cell was blocked with 5 mg/mL bovine serum albumin in PEM80 buffer supplemented with 20 mM dithiothreitol and 50 μM taxol. Dual-labeled microtubules (0.047 mg/mL) were introduced into the flow cell to bind kinesin (3 min). Excess/unbound microtubules were washed away with an anti-fade solution (20 mM dithiothreitol, 8.5 mg/mL glucose, 0.28 mg/mL glucose oxidase, and 210 mM catalase in PEM80 buffer) supplemented with 8 μM ATP. Streptavidin-fluorescein isothiocyanate (0.01 mg/mL in PEM80 buffer supplemented with 10 μM ATP, 20 μM taxol, and 0.2 mg/mL bovine serum albumin) was exchanged into the flow cell and incubated for 3 min. Excess streptavidin-fluorescein isothiocyanate was washed out with anti-fade solution. Finally, anti-fade solution supplemented with ATP (0-1 mM) and an ATP regenerating system[3] (2 mM phosphocreatine and 70 μg/mL creatine phosphokinase) were added to the flow cell to initiate kinesin-based microtubule gliding. The flow cell was then sealed using vacuum grease (Dow Corning, Midland, MI, USA) to prevent drying.

**Imaging and analysis**

Self-assembly of microtubules was imaged via epifluorescence microscopy using a fluorescence microscope (DM 2500P, Leica Microsystems Inc., Buffalo Grove, IL, USA) equipped with a Retgia Exi camera (QImaging, Surrey, BC, Canada). For each flow cell, five areas were observed; each area was typically imaged for 10.6 min at 40x magnification and at 0.17 Hz (0.5-1 s exposure).

Image analysis was carried out using Image Processing and Analysis in Java (ImageJ, http://imagej.nih.gov/ij/). The gliding velocity of microtubules was measured using the MTrackJ plugin for ImageJ (http://www.imagescience.org/meijering/software/mtrackj/). Microtubules typically assembled into linear bundles prior to spool formation.[4, 5] We tracked the trailing end of each microtubule bundle to determine the microtubule gliding velocity at each ATP concentration. The number of microtubule spools in an image was counted manually. The circumference of microtubule spools in an image was determined manually using the circle tool and the segmented line tool in ImageJ.

**Supporting Movie**

**Movie S1.** A representative movie of microtubule spool assembly at 1 mM ATP. The movie is sped up 5-fold.

Tan et al.

**Supporting Figure**

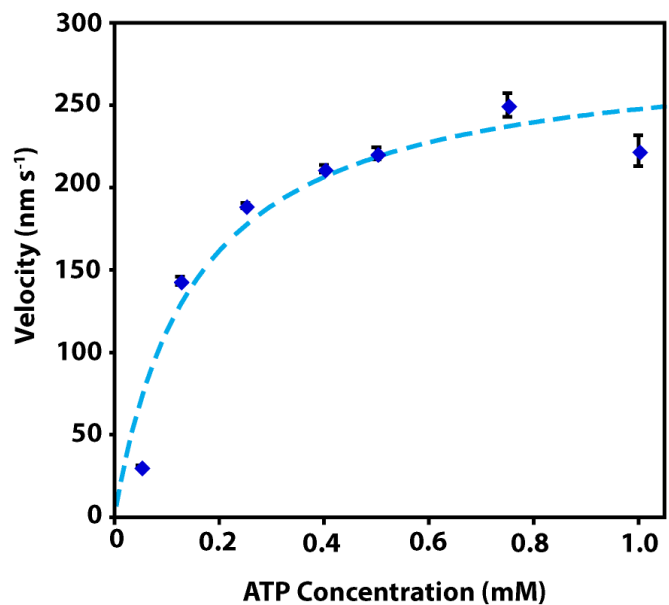

**Figure S1.**   Microtubule gliding velocity as a function of ATP concentration. The dependence of microtubule gliding velocity on ATP concentration is well characterized by Michaelis-Menten kinetics. Error bar, standard error of the mean. Dashed line, Michaelis-Menten kinetics ($K_m$ = 154±54 µM, and $V_{max}$ = 286±28 nm/s). At 1 mM ATP, the velocity of microtubule gliding (~200 nm/s) is consistent with the 100 nm/s reported previously.[6]

# Supporting References